\documentclass{sig-alternate-05-2015}

\usepackage{graphicx}
\usepackage{color}
\usepackage{balance}
\usepackage{booktabs} 
\usepackage{stackrel}
\usepackage{siunitx}
\usepackage{xspace}   
\usepackage{amsmath}
\usepackage{enumitem} 
\usepackage{url}
\usepackage{cite}
\usepackage{authblk}
\newcommand{\Paragraph}[1]{\smallskip\noindent{\bf #1.}}
\def\honeypot{SIPHON\xspace}

\begin{document}
\makeatletter
\def\@copyrightspace{\relax}
\makeatother

\doi{XXXX}

\isbn{XXXX}


%

\title{IoT Honeypot : Studying vulnerabilities of IoT devices}
\title{Honeytubes: Towards Distributed High-Interaction Physical IoT Honeypots}
\title{\honeypot: Towards Scalable High-Interaction Physical Honeypots}

%
%
%
%
%
\author[*]{Juan Guarnizo}
\author[*]{Amit Tambe}
\author[*]{Suman Sankar Bhunia}
\author[*]{Mart{\'i}n Ochoa}
\author[*]{Nils Tippenhauer}
\author[**]{Asaf Shabtai}
\author[*]{Yuval Elovici}
\affil[ ]{\textit {\{juan\_guarnizo, amit\_tambe, suman\_bhunia, martin\_ochoa, nils\_tippenhauer, yuval\_elovici\}@sutd.edu.sg, shabtaia@bgumail.bgu.ac.il}}

\renewcommand\Authands{ and }
\renewcommand\Authfont{\large}
\renewcommand\Affilfont{\large}

\maketitle

\begin{abstract}
In recent years, the emerging Internet-of-Things (IoT) has led to rising concerns about the security of networked embedded devices.  
In this work, we focus on the adaptation of Honeypots 
for improving the security of IoTs. 
Low-interaction honeypots are used so far in the context of IoT. 
Such honeypots are limited and easily detectable, and thus, there is a need to find ways how to develop high-interaction, reliable, IoT honeypots that will attract skilled attackers.
In this work, we propose the SIPHON architecture---a Scalable high-Interaction Honeypot platform for IoT devices. Our architecture leverages IoT devices that are physically at one location and are connected to the Internet through so-called \emph{wormholes} distributed around the world. The resulting architecture allows exposing few physical devices over a large number of geographically distributed IP addresses. We demonstrate the proposed architecture in a large scale experiment with 39 wormhole instances in 16 cities in 9 countries. Based on this setup, six physical IP cameras, one NVR and one IP printer are presented as 85 real IoT devices on the Internet, 
attracting a daily traffic of 700MB for a period of two months. A preliminary analysis of the collected traffic indicates that devices in some cities attracted significantly more traffic than others (ranging from 600 000 incoming TCP connections for the most popular destination to less than 50 000 for the least popular). We recorded over 400 brute-force login attempts to the web-interface of our devices using a total of 1826 distinct credentials, from which 11 attempts were successful. Moreover, we noted login attempts to Telnet and SSH ports some of which used credentials found in the recently disclosed Mirai malware.
\end{abstract}

%
%


%
%

%
%



\section{Introduction}

The Internet of Things (IoT) has gained immense popularity, creating a bridge between the physical world and the Internet.
According to the International Telecommunications Union (ITU) specifications, IoT can be defined as Internet connected physical objects or sensors which may be called as \emph{things}~\cite{itureport}. This implies that, unlike conventional network nodes in the Internet such as servers or PCs, IoT devices have a stronger link to physical reality by means of their sensors, and their behavior relies on that interaction with the analog world.
During the last decade, rapid advancement in sensor fabrication and miniaturization has fueled the growth of the \emph{things} tremendously.  It is expected that by the end of 2016, 6.4  billion \emph{things} will be connected to the Internet, growing to 11.8 billion by 2018~\cite{gartner2016}.

IoT has propelled the usage of applications such as smart wearable devices, intelligent transportation, and industrial automation (among others).
As a large number of \emph{things} are being used or 
deployed commercially, it is becoming challenging to 
provide suitable security mechanisms. Incorrectly configured 
\emph{things} will provide opportunities for attackers 
to perform malicious activities, ranging from 
sabotage of critical infrastructure to privacy 
violations. It is estimated that by the year 2017, 90\% of 
organizations that install IoT devices in their 
premises will suffer from attacks to their back-end IT 
systems, stemming from vulnerabilities in 
IoT devices~\cite{idc}. 

In fact, recent Distributed Denial of Service attacks on core internet services such as DYN's Managed DNS infrastructure were performed 
partially using IoT devices (IP cameras) that had default or weak hard-coded passwords on Telnet and SSH interfaces~\cite{dynattack}. 

Therefore, there is a strong need to develop suitable and 
cost efficient methods to find vulnerabilities 
in IoT devices, in order to address them before attackers take advantage 
of them. In traditional IT security, Honeypots~\cite{provos2004virtual,alata2007lessons,pouget2005advantages} are commonly used to better understand the dynamic threat landscape without exposing critical assets. Usually honeypots attempt to mimic a certain interaction in a realistic way (such as a login shell), 
encouraging unsolicited connections and potentially encouraging attempts to perform (possibly unknown) attacks.

Within the context of IoT, it is challenging to 
realistically mimic the interaction of IoT devices. For 
instance, consider the case of IP cameras. In order to virtualize or simulate their behaviour in a realistic way, one would need 
not only to broadcast some video to an attacker, but 
also react faithfully to commands such as tilting the 
camera or zooming in. Although not impossible, this 
requires a significant amount of technical work that 
cannot be easily reused for mimicking different types 
of IoT devices from different vendors, due to the heterogeneity of such devices.  

On the other hand, since IoT devices are sensing and actuating on the physical world, they might have a different perceived value to 
attackers depending on their location (e.g., different organizations or 
countries). This is similar to the trend seen on the botnet 
underground market where compromised servers in 
different regions are rented at different prices~\cite{holz2009learning}. Therefore a Honeypot exposing IoT devices should ideally appear geographically distributed to attackers interacting with it.

Based on the requirements mentioned above, a framework for deploying 
IoT honeypots should: (1) be able to provide \emph{high-interaction} in order to motivate skilled attackers to perform their activity and expose the vulnerabilities they are exploiting; (2) make it possible to easily ``place'' the IoT devices in a wide range of geographic locations; and (3) easily scale to devices of multiple vendors and of different kinds in order to keep track of a dynamic threat landscape.

In this paper we propose an abstract architecture 
and insights on a prototype implementation of such a Honeypot, that we call \honeypot. Our implementation allows us to deploy over 80 high-interactive devices with a diverse set of IPs located in different regions of the world by using only seven real IoT devices in our lab. We ran \honeypot for a period of two months gathering 20GB of raw traffic data. On doing a preliminary analysis of the data, we were able to see that instances in different cities received significantly different amounts of attention (in terms of connections and traffic). Moreover, curious users from the Internet attempted to brute-force the authentication of the devices, gaining access to the admin interface and interacting with devices in some cases.

In particular, we summarize the research problem and our contributions as follows.

\Paragraph{Problem Statement} IoT devices offer a rich set of potential attack vectors to the attacker. Our goal is to learn about existing and novel attack vectors on IoT devices. In particular, in this work we focus on gathering unsolicited traffic to IoT devices exposed to the Internet. Conventional honeypots use virtualization or simulation approaches to replicate the device under attack, and attract attackers. In the context of IoT, replicating the devices with virtual machines will be challenging. Only high interaction honeypots (e.g., allowing a user to move a camera) will convince skilled attackers to use advanced attack methods while exploiting zero-day vulnerabilities. The main question that motivates this work is \emph{How can we construct a large scale honeypot consisting of high-interaction IoT devices?}

\Paragraph{Approach} We propose the SIPHON architecture for scalable high-interaction IoT Honeypots. Our architecture uses physical IoT devices that are connected to various geographical locations through so-called ``wormholes''. The proposed architecture was implemented in a large scale distribution with 85 instances in 16 cities, nine countries while using only five physical IP cameras, one NVR and one IP printer (i.e., seven IoT devices). In our implementation, network traffic was forwarded leveraging an infrastructure of cloud service providers consisting of Amazon EC2 instances, DigitalOcean and Linode. We preliminary analyzed the collected traffic and found insights in terms of amount of interest in various geographical locations and types of interaction with the devices.

\Paragraph{Contributions}
The contributions of this paper are as follows:
\begin{enumerate}
\item We explain the expected benefits from a high-interaction IoT honeypot and define the design challenges.
\item We propose an architecture for a high-interactive IoT honeypot using limited number of physical IoT devices.
\item We present an implementation of that architecture based on seven devices presented as 85 distinct services on the Internet.
\item We analyze captured traffic, and show that geographic locations of our wormholes matter - attackers show more interest in some of the locations.
\end{enumerate}
 
We focus on high interaction distributed honeypot consisting of real IoT devices. That setup allows us to capture and analyze traffic and attacks across different geo-locations. To the best of our knowledge, we are the first to propose and implement such an architecture. 
 
This work is organized as follows. The background of this work is presented in Section~\ref{sec:background}. We propose the main architecture in Section~\ref{sec:approach}, and its implementation in Section~\ref{sec:implementation}. We analyze and discuss the traffic collected by our implementation in Section~\ref{validation}. Related work is summarized in Section~\ref{sec:relatedwork}. Finally, the paper is concluded in Section~\ref{sec:conclusions}.

\section{Background}
\label{sec:background}

A honeypot is an infrastructure commonly used to 
attract potential attackers to a controlled and monitored environment, without exposing critical assets, with the goal of understanding 
the threat landscape at a given point in time. Usually, it may consist of real or 
virtual systems~\cite{provos2004virtual} 
that mimic production environments~\cite{spitzner2003honeynet}. Interactions 
of the attackers with the honeypot are typically monitored and recorded for further analysis.
Researchers can then design appropriate mitigation techniques for  
attacks after analyzing the traces collected by the 
honeypot. From the point of intrusion analysts, ``honeypots provide another indication of wave of network attacks''~\cite{zhang2013security}. 

Honeypots may be set up in such a manner that they will 
provide convincing fake information which may be 
desirable to potential attackers, in order to attract their interest. 
Also, honeypots can 
help to redirect attackers to a decoy system and thus 
indirectly protect critical systems or infrastructure from a
possible compromise: if attackers are misled to think they are interacting with a real system, they are wasting 
time and resources to cause damage to real infrastructures. 

Security researchers classify honeypots as either high-in\-te\-ra\-ction or low-interaction in accordance with the privileges enjoyed by the attackers and real 
vulnerabilities being exposed to them. In the following we briefly recap these two categories.

\subsection{Low interaction honeypots}
Low-interaction honeypots emulate well-known vulnerable network services by partially implementing the TCP and IP stacks, in order to attract attackers. They do not give an attacker 
access to a real system, but rather to some sort of 
emulated system~\cite{fan2015taxonomy}. Since usually they do not run a fully featured operating system, they are considered to be safer in terms of remote 
exploitation. However, due to their simplicity, the main disadvantage of low-interaction honeypots is the higher likelihood of being detected as 
artificial by the attacker, especially if the attacker is a human. The attacker may find that 
the real services are emulated and services are only partially implemented, and thus lose interest. Low-interaction honeypots are by construction not optimal  
to capture zero-day vulnerabilities, since by their nature such vulnerabilities are unknown at the time of the deployment of the honeypot, and thus cannot 
be simulated~\cite{provos2007virtual}.
Low interaction honeypots can be used to detect a new wave of attack for a very limited time until the honeypot is labeled as such by attackers and dedicated search engines.

\subsection{High interaction honeypots}
High interaction honeypots help to observe attacks in a more realistic setting than low-interaction honeypots. They do not emulate any services, functionalities, or base operating systems. They thus allow to learn more about the tools, tactics, and motives of the attacker, because they do not restrict the attacker's behavior in the way that a low-interaction honeypot would~\cite{provos2007virtual}. 
As a result, high interaction honeypots are better suited to monitoring complex interactions between the attackers and the system~\cite{fan2015taxonomy}.

On the other hand, according to~\cite{fan2015taxonomy}, there are disadvantages to high-interaction honeypots. First they  are usually difficult and costly to 
deploy and maintain.  Also they pose a higher security risk, since attackers can stop normal operation of the device after scaling their privileges and may use 
the device at their will. If one is interested in exposing many vulnerabilities that might depend on different hardware/software configurations (like various OS versions etc.),  a large heterogeneous deployment of real devices is required. Last, although high-interaction honeypots cater for monitoring a wide range of events (such as network traffic and host-based activity such as system calls), this usually results in a large number of logs, making security analysis more challenging.

\subsection{Wormholes}
In the context of wireless communications, the term \emph{wormhole} refers to malicious forwarding of communication by the attacker~\cite{hu2006wormhole}, often through an out-of-band channel. In particular, wormholes can be used to influence routing schemes in dynamic sensor networks~\cite{hu2003packet}, or to attack wireless key schemes for cars~\cite{francillon2011relay}. Wormholes are typically very challenging or impossible to detect~\cite{hu2002wormhole}.

In this work, we will use the term wormhole in a novel, but related way. Instead of offensive wormholes used by an attacker, our wormholes are defensively used by 
the honeypot operator. To the attacker, the presence of the wormhole (and in particular the location of the other end) should be transparent. We discuss how to 
achieve that in Section~\ref{sec:approach}.

\subsection{Shodan}
\begin{figure}[htb]
\centering
\includegraphics[width=\linewidth]{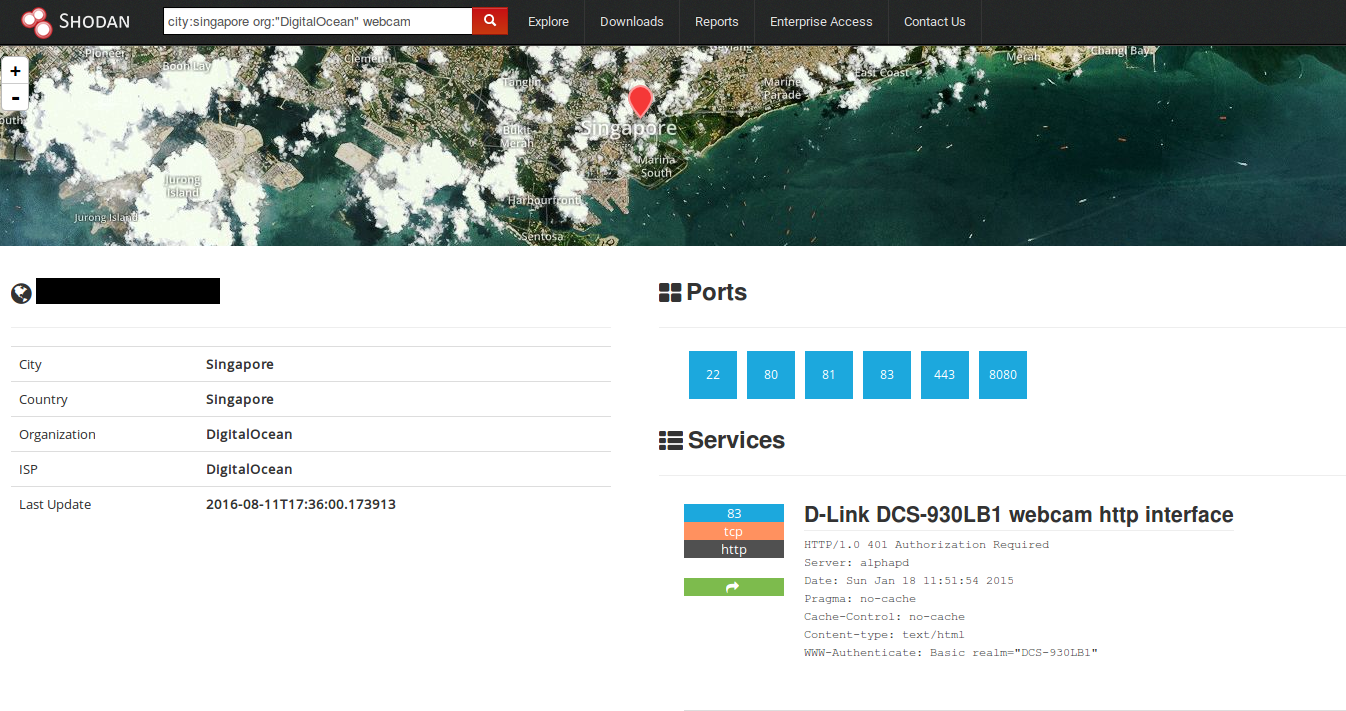}
\caption{Example of a camera listed on Shodan (in this case, one of our wormholes).}
\label{fig:shodan}
\end{figure}

Shodan \cite{bodenheim2014evaluation} is a search 
engine that lists IoT devices found on the internet, 
including Industrial Control Systems (ICS) devices such 
as Programmable Logical Controllers (PLCs) and household devices such as IP cameras among 
others. Figure~\ref{fig:shodan} shows a device in our Honeypot found on Shodan that corresponds to the search for \emph{cameras} in a specific location.

Shodan uses port scanning on internet IPs and analyzes 
the obtained responses to identify the type of device found. In 
average, it takes 1 to 2 weeks for a newly deployed 
device to be listed by Shodan~\cite{bodenheim2014evaluation}. 

In the following, our hypothesis is that Shodan will be 
used as the primary source of information on potential 
targets by attackers. Moreover, the Shodan API provides 
a service that tries to estimate if a device 
exposed by Shodan is a Honeypot or not. This service is called 
Honeyscore\footnote{\url{https://honeyscore.shodan.io/}
}, which is also accessible from the developer's command line interface API of Shodan. 
The Honeyscore uses a heuristic to classify devices as potential honeypots, giving a 
score between 0 (likely not a honeypot) and 1 (likely a honeypot).

\section{SIPHON: A Scalable High-In\-te\-rac\-tion Physical Honeypot}
\label{sec:approach}

In this section, we present our design for SIPHON: a Scalable high-Interaction Physical HONeypot. We start by outlining our attacker and system model, and then present the abstract system design.

\subsection{Problem Statement}
Our goal is to learn about existing and novel attack vectors for IoT devices. In particular we want to use the wisdom of the crowd of attackers in order to learn about the existence of vulnerabilities that are commonly exploited. In this work we focus on gathering unsolicited traffic to IoT devices exposed to the Internet. Conventional honeypots use virtualization or simulation approaches to replicate the device under attack, and attract attackers. In the context of IoT, replicating the devices with virtual machines will be challenging. Only high interaction honeypots (e.g., those allowing the user to move cameras) will allow to convince skilled attackers to use advanced attacks. The main question that motivates this work is \emph{How can we construct a large scale honeypot consisting of a limited number of real IoT devices?}

\subsection{Design Considerations}
In the following, we assume that attackers are using the Internet (and in particular Shodan~\cite{matherly2009shodan}) to identify potentially vulnerable or unsecured IoT devices such as webcams, smart fridges or similar. Typically, the attacker starts with a \emph{reconnaissance} phase, followed by an \emph{exploit} phase. The attacker could use either automated tools or manual interaction for either of those phases.

The devices under attack are assumed to be reachable from the Internet directly (i.e. on public IP addresses). They will expose one or more services on open ports. Such services could be HTTP, telnet, SSH, or a more specific protocol such as RTSP. The devices will be embedded in a real environment, i.e. the cameras will show live images, the fridge will contain real food, etc. Devices will be configured with minimal effort (i.e. potentially with default passwords). No dedicated security solutions are present (such as firewalls or IDS). If the devices are reachable through a NAT service, then the exposed services are forwarded by the NAT.

\subsection{Design Overview}
\begin{figure}[tb]
\centering
\includegraphics[width=\linewidth]{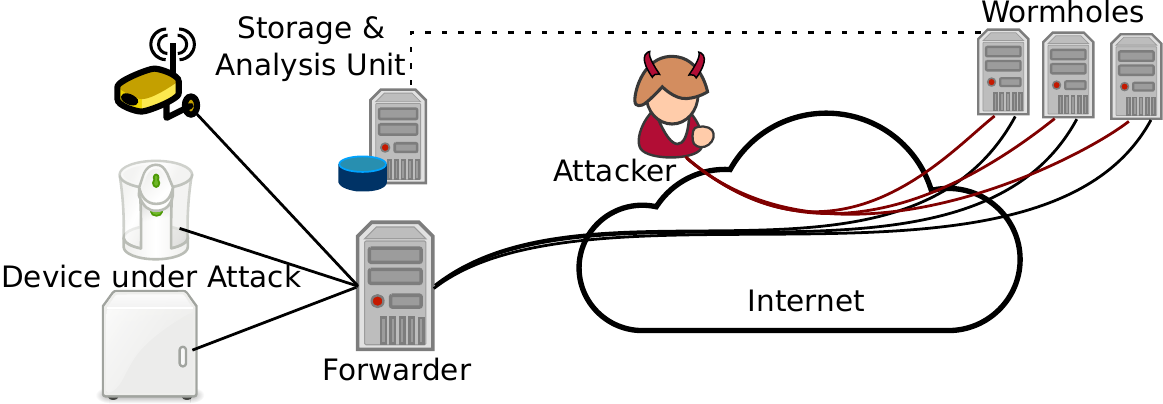}
\caption{Abstract overview of distributed physical honeypot}
\label{fig:overview}
\end{figure}

Our design (see Figure~\ref{fig:overview}) is based on real physical IoT devices, that are exposed to the Internet through a large number of \emph{tunnels} that forward traffic from \emph{wormholes} (remote public IP addresses) to the local physical IoT devices. 

\Paragraph{Wormhole} The wormhole device presents a number of open ports to the general Internet on a public IP address. Incoming traffic on those ports is either logged locally, or transparently forwarded to a specific port on a remote physical IoT device via the Forwarder device. There are a number of options on how this forwarding can be achieved, we discuss them in Section~\ref{sec:forwarding}. More than one physical device can be associated to each wormhole, but each service forwarded will need a unique port on the wormhole.

\Paragraph{Forwarder} The forwarder ensures that the traffic between the wormhole and the IoT device is re-written in real-time to hide the fact that devices are physically located somewhere else, or communicating with many other tunnels. 
If required, the forwarder can also act as a TLS man-in-the-middle because it will have access to the private keys of certificates used by the IoT devices. 

\Paragraph{Device under Attack (DUA)} The local (IoT) device under attack in the SIPHON architecture can be a normal commercial off-the shelf device. It will expose one or more services via the network, most likely over TCP. The device is able to maintain a number of concurrent connections at any point in time.

\Paragraph{Storage \& Analysis Unit (SAU)} The SAU obtains traffic records and general logs from the wormholes, and aggregates the data for offline analysis. 
For example, the recorded traffic can be analyzed by a suitable framework (e.g. Suricata or Snort~\cite{albin2011comparative}). In addition, dedicated analysis frameworks for other honeypot deployments could be used (similar to~\cite{yin2015IOTPOT}).

\subsection{Location of Wormholes}
As IoT devices are embedded in the physical world and interact with it, our hypothesis is that the physical location of the device under attack is correlated to its attractiveness to the attacker. While the direct physical location of a communication partner on the Internet is not easy to determine~\cite{androulaki14location}, many services on the web provide IP geolocation services (e.g.,~\cite{ipwhois}). 

For that reason, we propose to ensure that the wormholes have public IP addresses that are localized in a range of physical locations, ideally spanning multiple countries or even continents. 

That goal can be achieved in a number of ways. While nowadays the IPv4 address space is becoming scarce~\cite{richter15ipv4}, the easiest way to obtain public IP addresses seems to be a) home users (who often get one IP per subscription line), and b) servers on the cloud (cloud providers own large IPv4 address spaces). In this work, we focus on the latter (see Section~\ref{sec:implementation}). To take advantage of public IP addresses of home users, we also considered preparing devices such as Raspberry PIs with a specific configuration as wormhole to forward traffic to the honeypot forwarder. The disadvantage of that approach is the fact that such home setups would likely require the home user to set up port forwarding on the home NAT device (often called \emph{router} in that context). That configuration effort might be non-trivial for many users, and thus we have not followed this implementation path.
	
\subsection{Forwarding Method}
\label{sec:forwarding}
The tunnels between the forwarder and the wormholes can be realized using a range of different forwarding techniques. We now discuss a number of options that we considered.

\Paragraph{Application-Layer Proxy} If the honeypot should represent only few selected services, it would be possible to use proxies on the wormholes for those services (E.g., an HTTP proxy like mitmproxy). There are a number of advantages of that approach, for example the option to cache static information and thus reduce the traffic load on the DUAs.

\Paragraph{Encapsulation} On the transport and application layers, services such as SSH and TLS can be used to encapsulate selected traffic with low configuration overhead. In our implementation, we chose an SSH tunnel as method to forward the traffic.

\Paragraph{Network-Layer VPN} A number of network-layer VPN services are available (e.g., IPSec), a detailed comparison of which is out of scope here (we refer to~\cite{kotuliak2011performance}). Such services might have an advantage over selective forwarding using SSH, as large numbers of ports could be forwarded with little configuration.

\subsection{Compromise Detection and Restoration}
As the SIPHON architecture exposes real devices to the attacker, it is possible for the attacker to fully compromise a device as part of an attack. Once compromised, the devices could be used to attack other devices, the general neighboring infrastructure, or distribute malware to other attackers. There are several options to mitigate or prevent such scenarios.

\Paragraph{Periodic Reset}
A simplistic approach to mitigate the impact of compromised devices would be a setup that periodically resets the configuration and firmware of all IoT devices that are used in the lab. That approach has several disadvantages (i.e. timeliness, completeness of reset). On the other hand, it is easy to set up with some manual intervention.

\Paragraph{IPS} Intrusion prevention systems like the open source Suricata~\cite{albin2011comparative} could be used together with custom rules tailored towards the IoT devices to prevent their ongoing compromise. In addition, an IDS/IPS could be used to detect the changed behavior of a compromised device to identify that such a compromise has happened.

\Paragraph{Full instrumentation} The most reliable detection of a compromise could be achieved through low-level instrumentation of the IoT devices, e.g. through JTAG or similar connections. The setup and maintenance of such instrumentation is possible, but expected to be challenging and expensive~\cite{guri15joker}.

\subsection{Scalability}
As we claim in the acronym SIPHON, the proposed architecture should be scalable. Our intuition for that is as follows: for $n$ physical IoT devices and $m$ wormholes, the number of  devices presented to the attacker, without exposing the same device twice in the same wormhole, can be up to $n*m$. For instance, if $n=100$ and $m=1000$ up to 100,000 services could be exposed to the Internet. The number of services per wormhole is limited to around 65,000 due to port numbering restrictions. The number of wormholes per physical device is limited by the amount of parallel traffic that can be handled by each physical device.  In our practical implementation, we chose lower values for both $n$ and $m$ due to budget and space constraints as we will discuss in the following.

\section{A distributed IP-Camera Honeypot}
\label{sec:implementation}

In this section, we discuss an implementation of the proposed \honeypot architecture for a specific class of IoT devices: IP cameras used for home or public surveillance. We chose that class of devices as the number of IP cameras connected to the Internet is increasing daily, and recent research findings have shown that the IP cameras and Network Video Recorder (NVR) are vulnerable to attacks~\cite{yin2015IOTPOT, dynattack},
 thus attracting more unsolicited traffic than other IoT devices in the market. In addition, IP cameras allow access to private video, and often allow complex interactions with the user (e.g. manual movement control).
\begin{figure*}[!htb]
\centering
\includegraphics[scale=0.4]{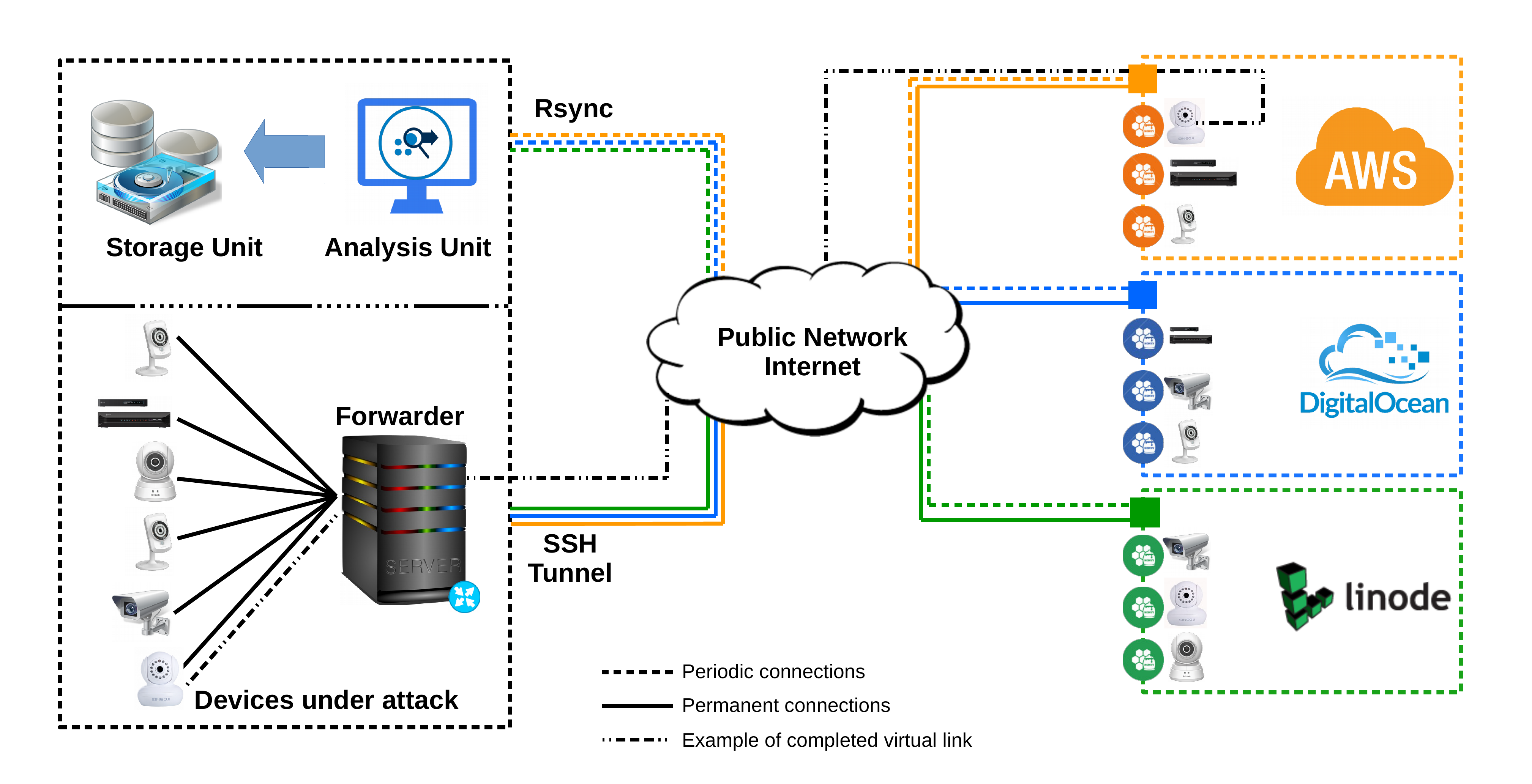}
\caption{SIPHON prototype implementation in our lab}
\label{infrastructure}
\end{figure*}



In our implementation, locations of the cameras are spoofed using wormholes based on instances in cloud services. The instances are deployed in different cities around the world. From the perspective of the attacker, the cameras are located across cities (e.g., by using the location feature of search engines such as \emph{Shodan}). All traffic is directed from the wormholes to our lab for in-depth analysis.

\subsection{Implementation}
Figure~\ref{infrastructure} shows our prototype infrastructure for the SIPHON deployment with IP cameras and a networked video recorder (NVR). Table~\ref{tab:device_details} shows the list of devices used.



\begin{table}[htb]
\begin{center}
\begin{tabular}{p{3cm}p{2cm}p{1.5cm}}
\toprule
\textbf{Device Model}       & \textbf{Password} & \textbf{Difficulty} 
\\ \midrule
HP Pro Printer 6830        & 1234567890        & Easy             \\ 
D-Link DCS-9050            & password123       & Easy             \\ 
D-Link DCS-930L            & YAQvwrjy          & Hard             \\ 
D-Link DCS-942L            & 1234567890        & Easy             \\ 
Aztech WIPC409HD           & admin             & Default          \\ 
Sineoji PT603V             & 9WgnTMxe          & Hard             \\ 
Trendnet Emulator          & admin             & Easy             \\ 
HikVision NVR 7604NI-E1/4P & xDk2PKHU          & Hard             \\\bottomrule 
\end{tabular}
\end{center}
\caption{IoT Device Details}
\label{tab:device_details}
\end{table}

\begin{figure}[h]
\centering
\includegraphics[width=\linewidth]{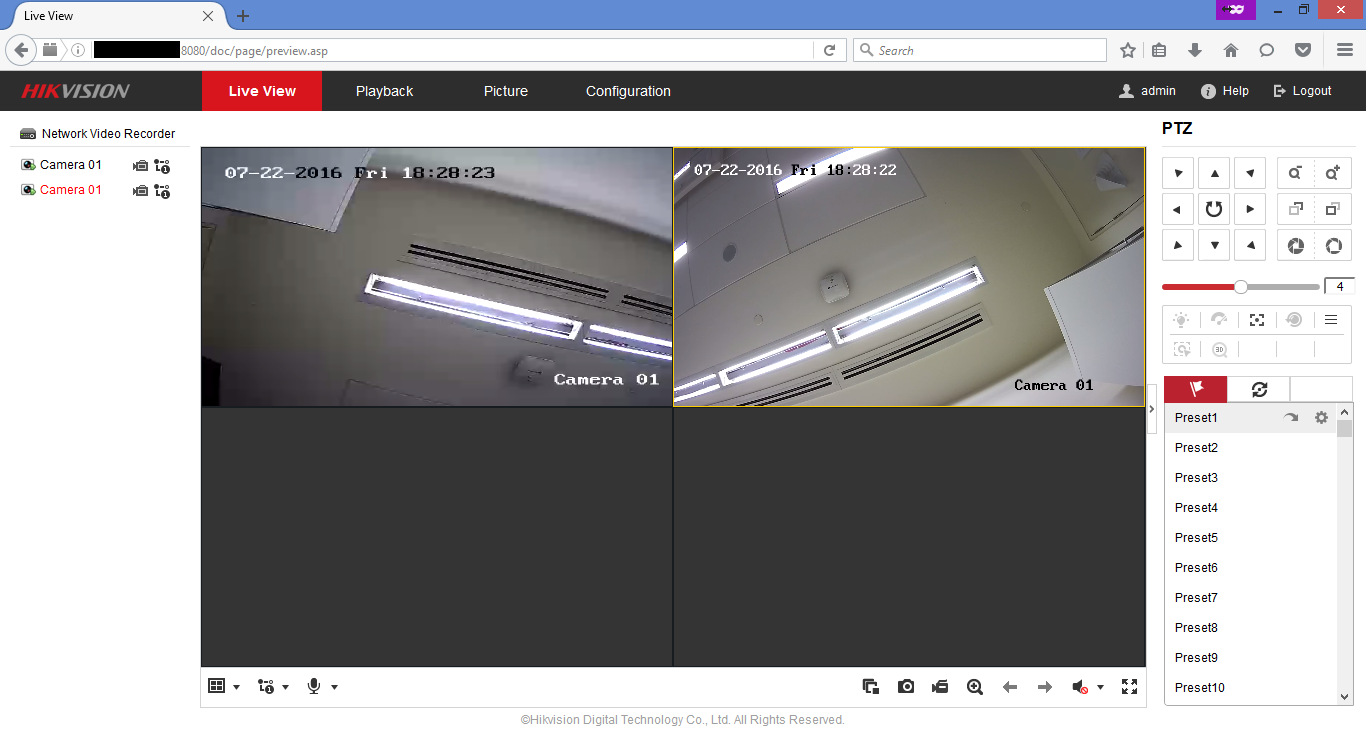}
\caption{Example of device view through a wormhole }
\label{fig:screenshot}
\end{figure}

By default, all the devices expose an administration web interface protected by a password. In order to attract complex interactions with the devices, we use default passwords for some of the cameras while we use weak passwords for others.
A screenshot of one of the HTTP interfaces of the cameras after successful authentication is shown in Figure~\ref{fig:screenshot}.


Using the administration web interfaces aligns well with our goal of capturing high interaction traffic. They provide apt opportunities to expose to the attacker, multiple device interaction scenarios, such as camera motion, wi-fi network scanning, etc. Even though more recent attacks, like the one using Mirai malware, rely completely on the discovery and exploitation of vulnerabilities via the Telnet and SSH ports~\cite{mirai}, we do not actively seek information on attacks made on these ports. This is due to the following reasons: 1) Telnet/SSH would not provide a high-interaction environment as an environment that enables video transmission  via HTTP and therefore would be less interesting 2) Telnet/SSH was already explored in this context by IoTPOT~\cite{yin2015IOTPOT} 3) None of the IoT devices used by us have open Telnet or SSH ports

\begin{figure}[h]
\centering
\includegraphics[width=0.9\linewidth]{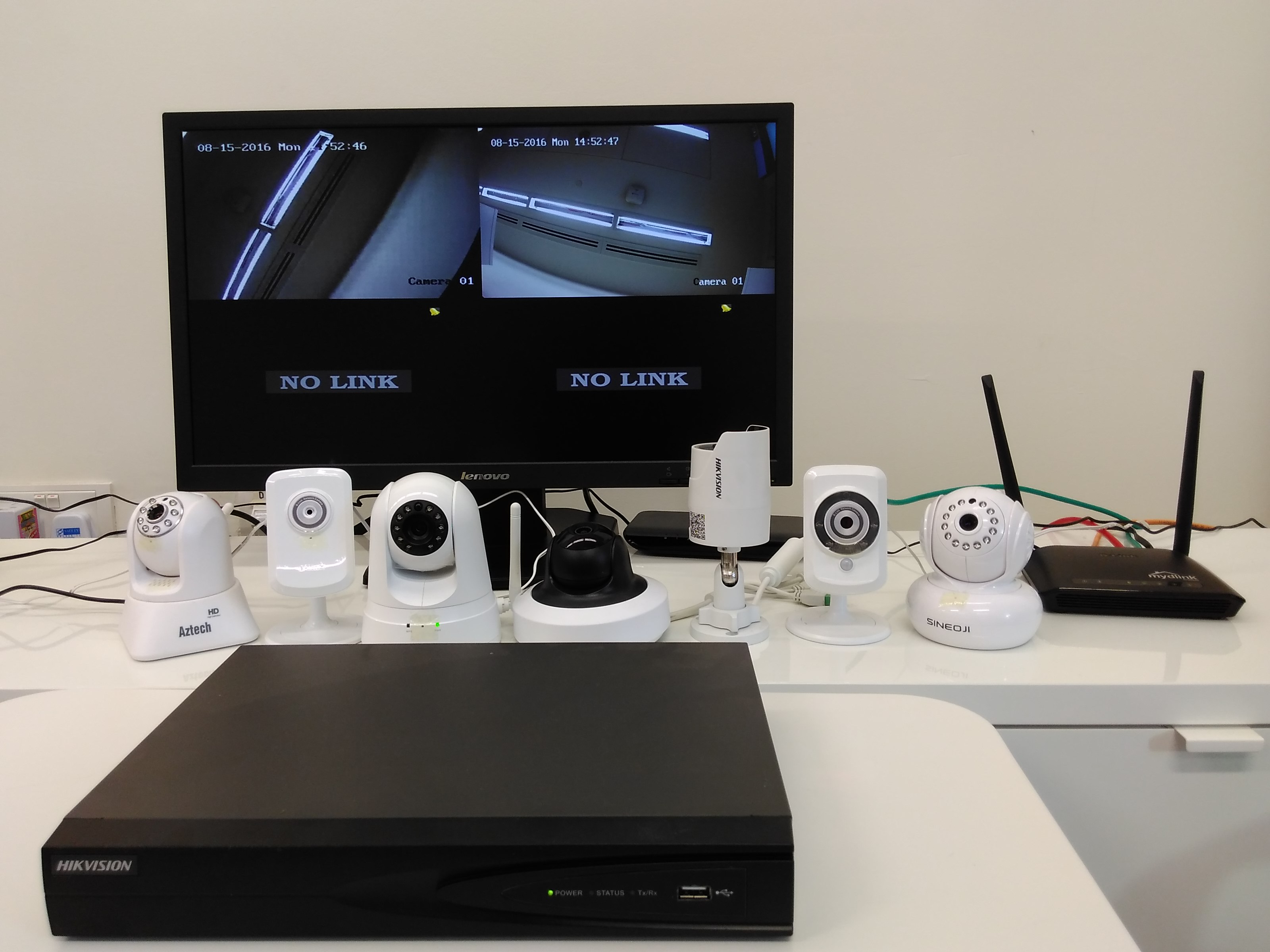}
\caption{Setup of Cameras and NVR in Lab.}
\label{fig:setup}
\end{figure}

The devices in our lab were connected to cloud servers from three different cloud service providers (viz. Amazon, LiNode and Digital Ocean) in various cities. During our experiment, a total of thirty nine servers were deployed and each of them had a distinct public static IP. Each of the cloud servers was designated as a wormhole with up to three (3) connected devices. The devices were connected on different ports of the cloud server. In total, 85 devices were visible to the attackers on the Internet, while only eight physical devices were used, thereby yielding in average about $10$ possible virtual connections to each physical device. \emph{Shodan} obtained the information about the devices using device fingerprinting heuristics such as parsing the HTTP response of the device and published the devices along with their geo-locations, accessible ports and other useful information as illustrated in Figure~\ref{fig:shodan}.

Eventually, all the traffic (i.e. device solicitations or attacks) was being diverted to the IP cameras or Network Video Recorders (NVR) in our lab. All traffic was captured using \texttt{tcpdump} on each of the wormhole servers and the resulting traces (in .pcap format) were stored locally for offline analysis.

\subsection{Technical Setup for Traffic Forwarding}

Our technical setup relies on a network gateway scheme, as described abstractly in Figure~\ref{fig:overview}. To implement the \emph{forwarder}, we used a local server in our lab which manages the TCP connections from the wormhole instances in the cloud. This server alone handles all $m$ wormhole to $n$ IoT devices communications. The server itself is a VM guest with 1 virtual core and 4GB of RAM, running Ubuntu 16.04.

The connection between the forwarder and the cloud instances is established with reverse \texttt{ssh} tunnels that redirect traffic of a specific port (i.e. port 80) on the wormhole to a port in the forwarder. Once the traffic has reached the forwarder, we complete the traffic redirection to the IoT devices by means of the \texttt{socat} linux command, which also forwards the device's responses to the cloud instance through the forwarder. This process is easy to automatize by using a different local port in the forwarder for each cloud replica of a physical device (wormhole).

The IoT devices are isolated from other devices in the lab through the use of 802.1Q VLANs. That ensures that even compromised devices can at most interact with other IoT devices, and the server.

\subsection{Locations of Wormholes}
For the attackers, the IoT devices are placed in the cities where the actual wormhole is placed. Physically, the IoT devices can be located in a common area together with the forwarder.

\begin{figure}[h]
\centering
\includegraphics[width=1.0\linewidth]{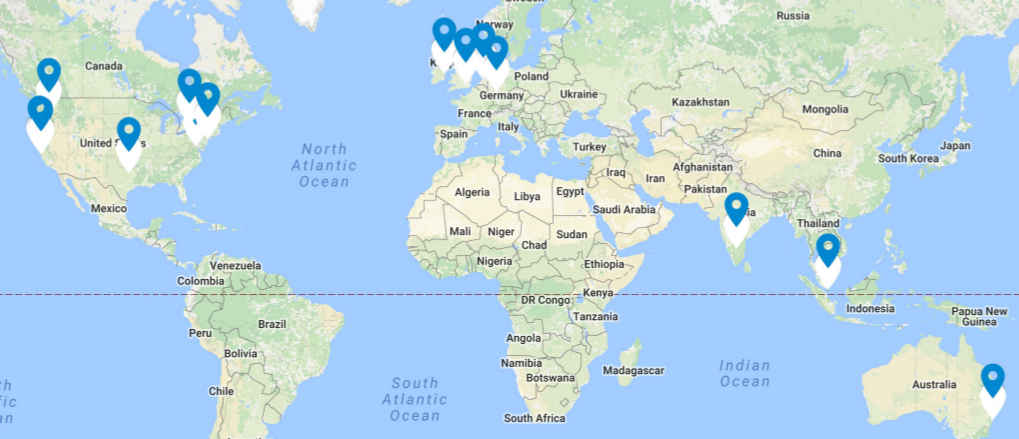}
\caption{Prototype wormhole locations in cities around the world.}
\label{map}
\end{figure}

Figure~\ref{map} shows the locations of honeypot deployment in various cities in different continents around the world. The \honeypot experimental deployment covered the following 16 cities in 9 different countries: 
a) USA (San Jose, Boardman, Ashburn, New York, San Francisco, Dallas, Fremont, Newark)
b) Canada (Toronto)
c) Europe (Frankfurt, London, Dublin, Amsterdam)
d) Asia (Singapore, Bangalore) 
e) Australia (Sydney).

This setup had a total cost (for 39 wormhole instances at three different providers) of about 12 USD per day. We note that the smallest instance sizes are usually sufficient for the setup, which allows to keep the cost low. 

\subsection{Hiding the honeypot character}
 The literature~\cite{provos2007virtual} defines different detection methods deployed by attackers to detect the existence of honeypots. Such methods usually vary depending on the tool used to create the honeypot. For e.g. a honeypot created using the tool `Nepenthes' can be detected using NMap (a network open port mapping tool). NMap is able to identify the version of only one service amongst multiple services that are running on a Nepenthes machine. On the other hand, a honeypot created using `Sebek' can be detected by discovering the memory addresses of the system calls `sys\_read' and `sys\_write' and ensuring that they are not more than 4096 bytes apart from each other. Usually, low interaction honeypots use tools that partially mimic network stacks (TCP/IP) to attract attackers by exposing vulnerable services. Such network stacks may be susceptible to well crafted packets thereby revealing the existence of a honeypot. Also, by their very nature honeypots built using virtual machines are subject to timing attacks. For e.g. the timestamp in TCP packets over different virtual machines hosted in the same physical machine can exhibit a similar skew over a period of time in contrast to the same setup over different physical machines~\cite{provos2007virtual}.

Although we have not yet fully explored ad-hoc attacks against our infrastructure, after running our distributed honeypot for a while, we checked whether its IP addresses are marked as ``honeypot'' by Shodan. During the experiment run, our IP addresses were mostly not detected as honeypot by Shodan (see Figure~\ref{fig:shodan}), in particular the score we get from Honeyscore through the Shodan's developer API ranges in most cases from $0.0$ to $0.5$ (not honeypot), with an average score of $0.16$ and only one instance was labelled as a Honeypot with confidence $0.8$. The exact method how Shodan is detecting honeypots is unknown to us, but most likely it is a heuristic based on well-known open source honeypot tools. 

\section{Preliminary Analysis of Traffic}
\label{validation}
In this section, we present the results of our analysis of the traffic captured by our prototype. In particular, we focus on the following aspects: a) whether the location of the wormhole matters to the attacker, b) whether some devices were more attractive to attackers than others c) whether being listed in Shodan had an impact in terms of unsolicited connections d) whether attackers would show different behaviour when interacting with a low-interaction version of our setup and e) the kind of interactions between the attacker and the devices.

\subsection{Analysis Goals}
Interests of attackers may vary depending on several factors as discussed above. In this work, the following hypotheses have been considered in this regard. 

\Paragraph{Location}
We conjectured that attackers prefer to invest more time in certain cities to ascertain their targets than others. They search for devices in a particular city of their interest and initiate interactions subsequently. This can be motivated for instance by a commercial interest of the location's IP for re-selling after infection (such as in the case of botnets~\cite{holz2009learning}), or as a starting point for targeted attacks.

\Paragraph{Device type}  
We assumed that attackers target particular models or device types that might have known vulnerabilities. For example, attackers may particularly look for vulnerable IP cameras. 

\Paragraph{Shodan Listing}  
We speculated that attackers may target devices more after they get listed on Shodan. We look for differences in the connection attempts before and after a device gets listed on Shodan. 

\Paragraph{High-interaction vs. Low-interaction} We conjecture that attackers might behave differently when exposed to a low interaction implementation 
of our devices, for instance a camera administration web-service that has an identical look and feel as the original, but shows a static image instead 
of video.

In order to validate the points above, we take into account the following interaction factors. First, we want to understand how many TCP connections we are receiving per wormhole, and what kind of services are being consulted (SSH, our HTTP ports etc.) This already sheds light on how many wormholes have been discovered and how much attention do they receive. 

Next we are interested in understanding whether such interactions are being performed using well-known scripts based on the number of connections received per wormhole in short time frames, the number of different ports accessed by the same wormhole, as well as the user agent in HTTP sessions (for example, in the collected traffic we identified the \emph{masscan} agent~\cite{masscan}).
Moreover, since to have access to our devices' admin interface an attacker needs to authenticate, we are interested in counting brute-force login attempts, to count how many of those have been successful, and to further investigate what happens after a successful login. 

To further evaluate attacker attempts to gain access to our devices, we have assigned default, easy and hard passwords (Table~\ref{tab:device_details}). We can then, by analysing the HTTP responses issued after successful login, gather statistics on movements or zoom-ins on the cameras, scanning for WiFi networks (a feature often offered in the admin interfaces of IP cameras) and more interestingly, firmware updates.\par

\subsection{High-interaction IP camera honeypot}

All the network traffic is collected and stored on a local machine in our lab. We store the raw pcaps obtained with \texttt{tcpdump}, and then in order to perform the aforementioned analysis, we parse the pcaps and store the basic features of each TCP connection in an SQL database.

\Paragraph{Distinct remote IPs} Upon analyzing the more than two months' worth of data, we observed incoming TCP connections from over 13000 distinct remote IP addresses.

\begin{figure}[h]
	\centering
	\includegraphics[width=\linewidth]{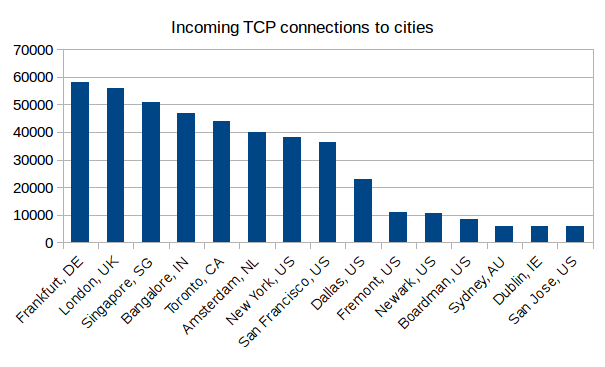}
	\caption{Distribution of TCP connections to wormholes according to their city}
	\label{fig:hp1_cities}
\end{figure}

\Paragraph{Connections to wormhole per city} Based on the geographical distribution of our wormholes, we observed that wormholes in some cities received more attention than wormholes in other cities. Figure~\ref{fig:hp1_cities} depicts the distribution of incoming TCP connections received by wormholes per city. The wormhole(s) in Frankfurt received the most connections (almost 600 000), while San Jose in US received the least number of connections (about 50 000). 

\begin{figure}[h]
	\centering
	\includegraphics[width=\linewidth]{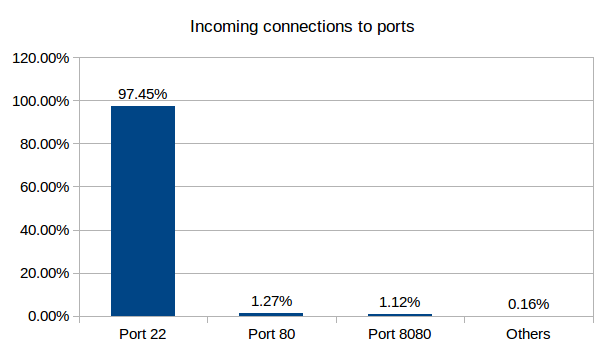}
	\caption{Connections to ports of wormholes}
	\label{fig:hp1_prt}
\end{figure}

\Paragraph{Wormhole ports} However, we observed that most of the incoming connections (about 97\%) were on port 22 (SSH) of the wormholes, whereas HTTP ports like port 80 and port 8080 received just 1.27\% and 1.12\% of total connections respectively. Other HTTP ports which we had opened got a mere 0.25\% of the total incoming TCP connections. Figure~\ref{fig:hp1_prt} shows the distribution of incoming TCP connections according to the wormhole ports.

\begin{figure}[h]
	\centering
	\includegraphics[width=\linewidth]{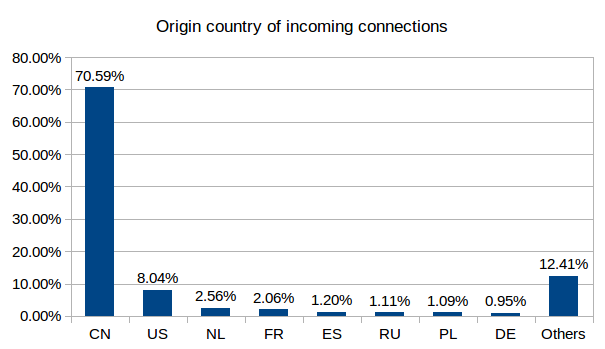}
	\caption{Incoming connections according to their country of origin}
	\label{fig:hp1_origin}
\end{figure}

\Paragraph{Location of attackers} Although not a reliable indicator (since attackers can easily tunnel through IPs located anywhere in the world), we also gathered statistics on the location of remote IPs. 
We analyzed the origin of all the incoming connections by country. Figure~\ref{fig:hp1_origin} shows the distribution of incoming TCP connections by their countries of origin. We saw that more than 70\% of the connections originated in  China followed by 8\% connections originating in USA. Netherlands, France and other countries make up the remaining list as shown in Figure~\ref{fig:hp1_origin}.

\begin{figure}[h]
	\centering
	\includegraphics[width=\linewidth]{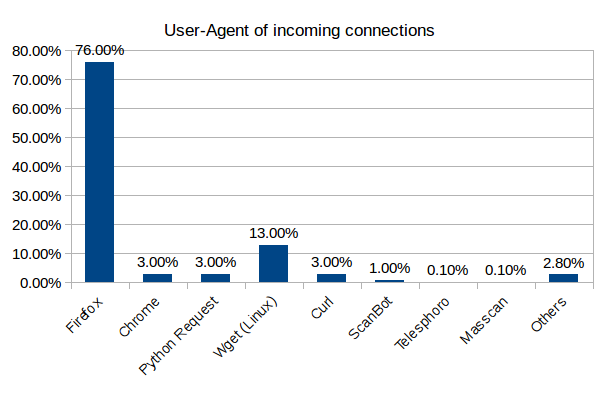}
	\caption{Incoming connections according to user agents}
	\label{fig:hp1_agents}
\end{figure}

\Paragraph{User-agents used by attackers} Next, we enumerated the agents being used in the connections to our devices. In Figure~\ref{fig:hp1_agents}, usage of different user agents in incoming TCP connections has been shown. Most of the connections (76\%) were made using Mozilla as the user agent. Apart from this, Chrome, Python Request, Wget(linux), Curl, Scanbot, Telesphoreo and Masscan~\cite{masscan} agents were used. Other user agents were combined together into the `Others' category that contributed 2.8\% of the total connections and included IP Camera Viewer, Cam Finder, Morfeus F Scanner, Msqq agents, etc. Usage of Nmap was detected several times. To top this, we also noticed around 9000 Shell shock attempts on our devices.

\begin{figure}[h]
\centering
\includegraphics[width=0.9\linewidth]{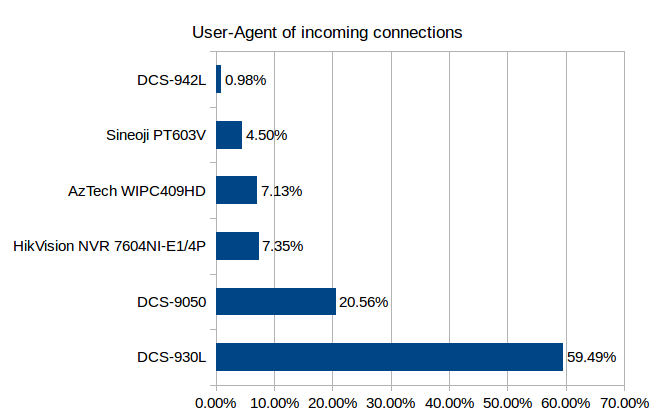}
\caption{Traffic breakup for the devices}
\label{fig:devices}
\end{figure}

\Paragraph{Attacked devices} In Figure~\ref{fig:devices}, we depict the combined amount of interest generated (based on HTTP traffic only) by the different physical devices. We observe that the majority of traffic was directed towards the DLink DCS-930L camera. We conjecture that this device attracted the most attention from attackers due to a recent report in the media about the presence of vulnerabilities in its firmware~\cite{dlink2016vulnerability}.

\Paragraph{Brute force login attempts} We have setup the IoT devices with different levels of password difficulties (Table~\ref{tab:device_details}). We retained the default credentials for the web interface login on some, while we modified the credentials on others. We configured some cameras to have easy to guess passwords, whereas others were configured with hard to guess passwords. With this setup, we observed 404 brute force attempts on all the devices combined from 137 distinct remote IPs. We deem a login attempt as a brute force attempt if the same remote IP tries to perform more than three login attempts in an HTTP session. Of these 404 brute force attempts, we observed a total of 11 attempts succeed. All the successful login attempts were on devices with easy passwords, and no login was recorded on a device with a hard password. We note that these easy passwords are present in most of the dictionaries used by automated brute-force attacks. On the other hand, these passwords can be guessed or brute forced easily even when attacking manually. 

\subsubsection*{Impact of device listing on Shodan}
Another goal of this analysis was to investigate the impact of the listing of a wormhole on Shodan. We tried to understand this by computing the number of TCP connections received by wormholes before and after their Shodan listings. Shodan usually takes between one-two weeks to enlist an IP after it becomes live on the Internet. With respect to this, we took into consideration one week's worth of incoming connections before and after Shodan listing. 

\begin{figure}[!htpb]
	\centering
	\includegraphics[width=\linewidth]{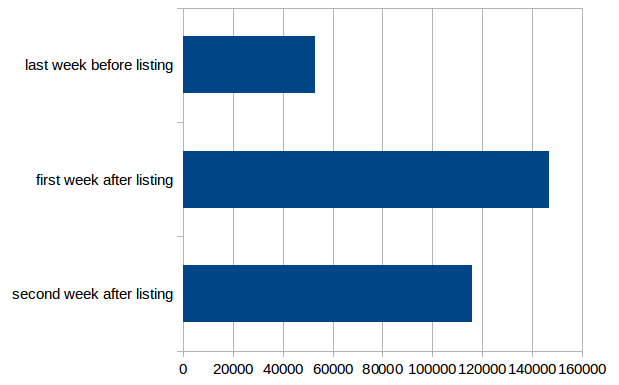}
	\caption{Impact of Shodan listing on no. of incoming connections}
	\label{fig:shodan_average_impact}
\end{figure}
In Figure~\ref{fig:shodan_average_impact}, we show the combined impact of Shodan listing for all wormholes. In particular, we show how many average connections the wormholes attracted per week, before and after the listing. In Figure~\ref{fig:shodan_average_impact}, the top bar represents the average number of connections across all the wormholes before they get listed on Shodan. The bar in the middle represents the average number of connections during the first week after getting listed on Shodan, while the bottom bar represents the average number of connections during the first two weeks after getting listed on Shodan. A surge in incoming connections is evident after listing on Shodan. We also observed that this surge in connections continues even in the second week after Shodan listing, but is reduced in increase. Thus, we can infer that there is an immediate impact of the Shodan listing, and attackers try to compromise the devices immediately after listing on Shodan. In the long term, the number of incoming connections decreases.


\subsection{Other IoT devices}
So far, we have used only IP-cameras and an NVR in the high interaction honeypot and gained knowledge about the attackers' behavior/interaction with them. But these devices do not represent the entire spectrum of existing IoT devices. We are interested in garnering knowledge about attackers' behavior towards other IoT devices as well. Therefore, we added another IoT device, the IP printer, to our honeypot, for about four weeks. This IoT device was connected to a cloud instance of Digital Ocean and located in London. Thus, similar to the IP cameras, even though the device was physically present in our lab, it appeared to be located in London. 

After analyzing four weeks' worth of network traffic data from the IP printer, we present our results. 

\Paragraph{Wormhole ports} We observed that most of the connections (99\%) were on port 22 (SSH) of the wormholes, whereas port 80 (HTTP) received a meager 1\%. 
Even in the case of IP cameras, we observed that the SSH port received more number of connections compared to HTTP ports.

\Paragraph{Location of attackers} Similar to the observation in the case of IP cameras, we noticed that the highest percentage of incoming connections to the IP printer (79\%), originated in China. Taiwan was next with 11\% of connections originating in Taiwan. France, India, Kazakhstan, Netherlands, Ukraine and USA made up the remaining portions.


\Paragraph{Brute force login attempts} We had changed the credentials of the IP printer to a password that should be easy to guess (\textit{admin:1234567890}). During our experiment, we observed only one brute force login attempt. Similar to the case of IP cameras, we deem a login attempt as a brute force attempt if the same remote IP tries to perform more than three login attempts in an HTTP session. A total of four successful login attempts were noted, from three distinct remote IPs, all from China.

\Paragraph{User-agents used by attackers} Similar to the case of IP cameras, we tried to understand which different user-agents were used in the connections to IP printer. We observed that all the four successful login attempts were made using the Mozilla user-agent. Unlike in the case of IP cameras, we did not notice a few user-agents (viz. \textit{cameraviewer, IPcamera finder, Scanbot, Morfeus F scanner, Msqq etc.}). In addition, we did not notice any \textit{Shell shock} attempts on the IP printer either.

\subsection{Low interaction honeypot}
One of our hypothesis for the low interaction honeypot is that most of the attacks are executed in an automated manner using bots. We assume that these bots are not capable to distinguish between real and virtual or fake devices. So, these bots may conduct repeated attempts considering the device as a real one. In contrast, a human being will discover a fake device and will not make repeated attempts consequently.

We have implemented a low interaction honeypot using the same methodologies as of the high interaction one. However, instead of a real physical device, we used the Trendnet camera emulator in the low interaction honeypot. Hence, this camera is a fake camera and not a real one. The entry point (i.e. wormhole) to the low interaction honeypot has been created using an instance from cloud service provider Linode, located in Singapore. Thereafter, we gathered data from the low interaction honeypot for almost six weeks. 

The results from six weeks of data are now presented.
\Paragraph{Wormhole ports} We observed that most of the connections to the fake camera i.e. Trendnet camera emulator (87\%) were on port 22 (SSH) whereas HTTP Ports 8080 and 80 received 12\% and 1\% of the total incoming connections respectively 
. In that respect, the behavior seen is consistent with the trends observed with the high interaction honeypots.

\Paragraph{User-agents used by attackers} Similar to the corresponding observation in the case of high interaction honeypot, we observed the maximum percentage of incoming connections to be having the user-agent as various versions of Mozilla. Apart from that, the attackers also used WGET (Linux), python, masscan and Probethenet Scanner. All these user agents were observed in the high interaction honeypot as well.

\Paragraph{Login attempts} 
We have noted three successful login attempts with the default credentials \textit{admin:admin} from three distinct remote IPs. One successful login attempt was made each from China, Russia and Iran. 

We also observed that the attackers did not spend enough time to explore the device or its functionality. They spent on an average only around 30 seconds in the low interaction honeypot, compared to about one minute or even up to an hour in the case of real cameras. The time spent by the attacker exploring the device is calculated based on the timestamps of the first and last packets in the TCP session captured via tcpdump. It is understood that the duration of a session is correlated to human intervention. \textit{session time} will be longer when the attacker will try to explore functionality of the device. Human beings can easily distinguish between a still image and a streaming video unlike automated bots. Thus, we can more easily determine manual attacks using this low interaction honeypot compared to our earlier high interaction honeypot.

\subsection{Discussion and comments on validity}
By analyzing the traffic, we see that our proposed architecture was 
successful in attracting a high number of connections, which confirms our overall design. In addition, we were able to observe 
interesting interactions in our preliminary analysis of the gathered data. We note that in our analysis, we assume that interaction between the attacker and IoT device over HTTP is more likely performed by humans than scripts, since interaction with unknown HTTP interfaces is harder to automate~\cite{yin2015IOTPOT}. Our prototype is thus a 
promising first step towards effectively learning the threat landscape in IoT security.

We are aware of several issues with our 
current prototype. First, one might argue that an 
intelligent attacker can spot our honeypot in 
different ways. For example, the attacker can detect that two wormholes represent the same physical device by simultaneously logging in to the same devices from two different wormholes. We acknowledge this shortcoming, but note that during our experiment, our data indicates that at least no single IP has discovered all the wormholes 
and the majority only interacted with a small subset (typically 5 or less). The issue can also be mitigated by having one different physical device per wormhole. Since we were interested in comparing different locations, we decided to scale the system in a 1 to $n$ fashion instead of a bijective relation between devices and wormholes.

Compared to related work such as~\cite{yin2015IOTPOT}, we have so far observed very few attempts to upgrade the firmware, which is a very common objective of attackers who want to spread botnets or malware. We conjecture a possible explanation for this as follows. We have noticed that only the minority of traffic in the wormholes was directed to the ports exposing the web interface of our IoT devices. Attackers seemed much more interested in SSH connections, possibly because common automated attack tools are tailored to re-flash firmware using a shell, which is a general approach that works in many devices. We believe that the firmware upgrade attempts we received were performed manually, but we also think such attacks will increase in the future as attack tools automatize this process for different brands. It is also likely that in the future access to SSH ports on many devices such as home-cameras will be more restricted, whereas access to the web-interface might be still desirable for convenience of access. 

Additionally, one might argue that an IoT device 
connected to the Internet through an IP that is owned by 
Amazon is suspicious to start with. However, given the 
amount of interest received in terms of traffic and the 
fact that there were some (potentially manual) logins, 
we think that using IPs belonging to cloud services 
might not be considered suspicious to a wide range of 
attackers. Indeed, the fact that most of our wormholes 
are classified as not Honeypots by Shodan's Honeyscore 
(score of $0.0$), preliminarily indicates that our 
setting appears to be realistic from a technical point of view.

Incidentally we notice at least other 6 cloud instances 
exposing cameras to Shodan. All of those instances had 
a Honeypot score of $0.5$, which in principle makes 
them more suspicious than our set-up, but indicates 
potentially that either such a setting is not uncommon 
(if not honeypot), or that other people have been less 
stealthy in their Honeypot setup than us.


\section{Related Work}
\label{sec:relatedwork}
Honeypots are a common measure to understand attacker activities in computer networks. The authors of~\cite{fan2015taxonomy}  provide a taxonomy of honeypots, and differentiate between low and high interaction honeypots. Both low and high interaction honeypots are compared in~\cite{alata2007lessons}, and the authors conclude that high interaction honeypots provide more insights on attacker behavior than low interaction ones.

The Honeynet Project~\cite{spitzner2003honeynet} is a well-established project that focuses on monitoring and analyzing attacks to complement intrusion detection tools. The Honeynet project does not use emulation, and instead leverages real systems and applications. For that reason, the Honeynet project is an example of high interaction honeypots.

In~\cite{grizzard2005}, the authors describe honeynets that can be used to increase security in a large computer network. In short, honeynets are clusters of honeypots. 

A low interaction honeypot system is implemented in~\cite{kim2004design}. The honeypot monitors behaviors and learns the advanced attacks that may not be detected by IDS tools. The session of the attacker is redirected to the honeypot system, which then serves requests from the attacker. For that purpose, the authors provided service daemons and a fake shell so that the attacker is not able to discover the system as honeypot.

In~\cite{kim2007decoyport}, the \emph{DecoyPort} system was proposed which redirects attackers towards honeypots. The system creates \emph{DecoyPorts} on active computers, those ports are not used by real services. Whenever any query or request comes to the ports, the system diverts the same to the honeypot. The system does not only act as a port forwarder but also is capable of controlling network loads caused by attackers. 
With this high interaction honeypot deployment, authors noted that possibility of attack may increase.

In~\cite{canali2013ndss}, the authors described the implementation and deployment of a honeypot based on a number of real,
vulnerable web applications. They hosted all the web applications
in seven isolated virtual machines running on a VMWare Server.
In order to limit the attack
surface, the authors let the exposed services run as a non privileged user. They then analyzed the collected data to 
study attackers' behavior on the web applications during pre and post exploitation. In contrast, we work with real physical IoT devices to set up a high-interaction honeypot.

Among the related work on general honeypots, we found only one that focuses on IoT~\cite{yin2015IOTPOT}. In that work, the authors present a low interaction honeypot for IP cameras and Digital Video Recorders (DVR). The authors emulated \emph{Telnet} services for those devices. No real devices were used to deploy the honeypot. The authors' goal was to capture telnet based attacks and analyze the same with respect to the concerned IoT devices.

Another interesting aspect has been described in \cite{pouget2005advantages}. The authors discuss the necessity of deploying distributed honeypots for monitoring local trends of attack and location specific attack behavior. The authors deployed low interaction honeypots in different countries with active participation from various academic, industrial and governmental organizations. These honeypots emulate various operating systems. The authors have shown the similarities and differences in attacks on honeypots in different locations. 

A game theoretic model has been proposed in~\cite{pibil12gametheory}. That paper models a scenario when an attacker decides which server to attack in a network. The authors describe a honeypot selection game which will lure attackers to attack a honeypot instead of an actual server. They studied and analyzed attacker behavior empirically. They have shown that in addition to honeypot design, the strategies should also be emphasized.

Another recent paper~\cite{tonyjemin2016} proposes a Bayesian game to defend against attacks in a honeypot enabled IoT network. In that paper, various game scenarios are described depending on the changing strategies of both the attacker and defender. The authors perform systematic mathematical analysis of the games and evaluate the Bayesian model through simulation.

Compared to existing literature, our work focuses on a high interaction distributed honeypot consisting of real IoT devices. That setup allows to capture and analyze real traffic and attacks across different geolocations and measure attacker behavior with respect to different locations. To the best of our knowledge, we are the first to propose and implement such an architecture.

\section{Conclusions}
\label{sec:conclusions}

In this paper, we have proposed a design for a distributed and scalable high-interaction IoT honeypot. This design allows portraying physical IoT devices in a single lab as being geographically distributed by establishing tunnels between the public IP addresses and the physical devices. It also allows to collect traffic for further processing and analysis. We have implemented a prototype of this design using five IP cameras, one NVR, making available 85 IoT devices to the attackers worldwide. Our implementation has allowed us to gather several GBs of traffic data that we have preliminary analyzed particularly from the point of view of locations with higher attractiveness to the attacker.

In the data gathered we have particularly noticed 
more than 400 brute force attempts to bypass the 
authentication of our devices that have led to 42 
successful logins into the devices. We have observed an overwhelming amount of traffic to our devices but comparatively not so many interactions with the device's web interface. We conjecture that this is due to little automation in attack tools to perform such attacks, and comparatively less amount of manual attacks compared to automated attacks. Among all the deployed 
devices, one particular model of DLink camera got 
around 60\% of the total traffic, which we conjecture 
is due to recently published vulnerabilities for this 
model. Devices in Frankfurt and Singapore, attracted 
the highest amount of traffic, confirming our 
hypothesis that \emph{location matters}.

We have inducted other IoT devices like IP printer into the high interaction honeypot and tried to understand attackers' interaction. Moreover, we also deployed a low interaction honeypot using IP camera emulators to compare attackers' behavior with that of a high interaction honeypot. We observed that attackers are more interested in real IP cameras than emulators or other devices. 

In the future, we plan to add more devices to our Honeypot, such as home appliances. We also plan to perform analysis in more depth with the collected data as well as to share our data with other researchers interested in performing such analyses. Apart from this, we shall incorporate intrusion prevention system to prevent substantial damage to our devices.
Besides, we plan to integrate the SIPHON framework with the IoT testbed presented in~\cite{siboni2016advanced}. By incorporating SIPHON with the IoT testbed, we aim at achieving the following two goals. First, we can use the SIPHON as one of the security testing mechanisms of the IoT testbed. This can be achieved by exposing any IoT device that in tested in the testbed to the Internet via the SIPHON framework as a honeypot, and by using the measurement (monitoring) modules in the testbed we can detect attacks originating from the Internet and potential weaknesses of the IoT device. Secondly, we can use the IoT testbed as part of the SIPHON mechanism. By placing an IoT device in the testbed, we can use the stimulation modules of the testbed for increasing the reliability of the SIPHON honeypot. For example, we can send various GPS locations using the GPS spoofer, simulate a network to which the IoT is connected or simulate user activity with the honeypot.  

\bibliographystyle{abbrv}
\bibliography{bibliography}
%
\end{document}